\documentstyle[aps,epsf]{revtex}
\begin{document}
\twocolumn[\hsize\textwidth\columnwidth\hsize\csname @twocolumnfalse\endcsname
\title{Disclination Dipoles as the Basic Structural
Elements of Dielectric Glasses}
\author{V.A. Osipov and S.E. Krasavin}
\address{
Joint Institute for Nuclear Research,\\
Bogoliubov Laboratory of Theoretical Physics\\
141980 Dubna, Moscow region, Russia\\
}
\address{\em (\today)}
\preprint
\draft
\maketitle

\begin{abstract}
We show that the experimentally observed behavior of thermal
conductivity of dielectric glasses over a wide temperature range
can be explained by a combination of two scattering processes.
The first one comes from the phonon scattering due to biaxial dipoles
of wedge disclinations while the second one is the
Rayleigh type scattering. The results obtained support
the cluster picture suggested earlier for glassy materials.
\end{abstract}
]

Physics of glassy systems has been attracting a considerable
interest for a long period.
Since the discovery of the anomalous low-temperature behavior
of amorphous dielectrics~\cite{zeller} there were many
attempts to explain this unusual phenomenon.
And yet, in 1986 Freeman and Anderson~\cite{freeman} reviewing
both numerous empirical results and theoretical models made
a rather drastic conclusion that the thermal conductivity
of amorphous dielectrics "is not understood in any temperature range".
While at first glance this conclusion looks too pessimistic,
it reflects an actual situation existed in glassy physics at that time.
It has been found experimentally that the low-temperature anomaly in
thermal conductivity, $\kappa$,
observed by Zeller and Pohl for some noncrystalline solids~\cite{zeller}
is universal.
In particular, a $T^2$ dependence of $\kappa$
below $1 K$ together with the following characteristic plateau region
have been found in the most of glassy materials
(see, e.g., review~\cite{JTP})
as well as in some quasicrystalline alloys~\cite{chernikov}.
This finding allows to assume that a realistic theoretical
model for description of the thermal conductivity should capture
the very essence of these materials, their microstructure.
Thus the conclusion of Freeman and Anderson actually suggests
that this problem still remains to be solved.

The most widely used model which explains glassy physics
below $1 K$ has been proposed by Phillips~\cite{phil1}
and Anderson, Halperin, and Varma~\cite{AHV}. In this temperature
region, thermal transport takes place via propagating acoustic
modes. Thus, $\kappa$ depends on the scattering of the phonons
by the structure. Within the model~\cite{phil1,AHV},
the principal scatterers were proposed to be the tunneling states (TS)
in the glass.
However, in spite of the success achieved in interpreting
experimental data, some important questions were left unanswered.
The main of them concern the nature of TS and the source
of their universality (see, e.g., discussion in~\cite{JTP,GKK}).

Since 1972 there have been many attempts to explain the
experimental data in terms of propagating thermal phonons
scattered by TS~\cite{JTP,zaitlin,smith,jones,maynard}
(in different modifications), density fluctuations (Rayleigh
scattering)~\cite{zeller,JTP,jones,morgan,walton,graebner}, and the like.
Some papers invoked nonpropagating vibrational modes as well~\cite{JTP}.
The most successful approach includes a combination of TS and the
Rayleigh scattering~\cite{jones}. Notice that
there were attempts to go beyond the TS approximation
(see Ref.~\cite{kuhn} and the references therein) as well as
to do without TS at all. For example,
an interesting phenomenological approach was discussed
in Ref.~\cite{JTP} where a combination of the structure scattering
together with a constant free path was shown to give a good
fit to the thermal conductivity data in amorphous arsenic (a-$As$).
As a result, it was suggested that no tunneling state scattering
is needed to explain the temperature dependence of $\kappa$ in this
material.

Of our special interest is the model proposed
by J.C. Phillips~\cite{JCP}, who suggested that glasses
have cluster structures. It has long been known
that there is a close connection between grain
boundaries in materials with pronounced cluster structure
and wedge disclination dipoles (WDD)~\cite{li1}.
Recently, this analogy has been successfully applied for
description of the old experiments in polycrystals
on the basis of the WDD-induced scattering~\cite{osipov}.
As was mentioned in~\cite{JTP}, the scattering
in polycrystals has a close parallel in amorphous solids.

Another approach which takes into account topological defects
in glasses came from geometrical and topological considerations.
For example, Kl\'eman (see review~\cite{kleman})
proposed that disordered covalent systems can be described
in terms of distributions of defects, most probably
disclinations, in a crystal situated in a space of constant
curvature. This idea was explored by Nelson~\cite{nelson}
in description of metal glasses.
Similarly, Duffy and Rivier~\cite{duffy}
considered a continuous random network formed by 5 and/or 7 rings
in the hexagonal lattice. In this case one can also expect
the appearance of 5-7 disclination pairs (dipoles of
disclinations). Let us mention also papers~\cite{grannan} where
elastic dipoles have been introduced within the glassy model.

In this Letter, we present a new look at the problem
of thermal conductivity in dielectric glasses as well as
at their microstructure. Actually, we reexamine
the old ideas of the previous two paragraphs from another point of
view. The model proposed here was inspired by a recent finding that
a biaxial WDD is a rather unique scatterer~\cite{osipov}.
Namely, it has been
shown that the mean free path $l_D$ of thermal phonons scattered by
biaxial WDD with nonskew axes of rotation behaves like
$l_D\sim q^{-1}$ at small $q$ while $l_D\rightarrow const$ with $q$
increasing.
As is known, a similar behavior, but with $l_q\sim q^{-2}$
for small $q$, has been found in old experiments on glasses~\cite{berman}.
However, the later experiments showed that in fact
$l_q\sim q^{-1}$ in agreement with the behavior of $l_D$.
The main aim of this paper is to show that the phonon scattering
due to biaxial WDD, being combined with the Rayleigh scattering, can
explain the thermal conductivity experiments in amorphous dielectrics
over a wide temperature range. As an example, we compare the
results obtained from the experimental data for vitreous silica
(a-$SiO_2$). Moreover, considering WDD as
the basic structural elements of glasses one can explain the
universal behavior seen in different amorphous dielectrics.

To gain a better understanding, let us discuss briefly
the generally accepted approaches as well as the main problems
pointed out in~\cite{freeman}.
The starting point for the calculations
is the known kinetic formula
\begin{equation}
\label{eq1}
\kappa = \frac{1}{3}\int_0^{\omega_D} C(\omega, T)vl(\omega, T)d\omega,
\end{equation}
where $C(\omega, T)d\omega$ is the specific heat contributed by phonons
within the frequency interval $d\omega$, $v$ is an average phonon
velocity, and $l(\omega, T)$ is the mean free path of a phonon.
As usual, at low temperatures we can assume a constant $v$ and a density
of phonon states quadratic in $\omega$ with the Debye cutoff
$\omega_D$. To reproduce the observed $\kappa$ over the full
temperature range the best form of the total mean free path
was taken to be~\cite{jones}
\begin{equation}
\label{eq2}
l(\omega, T) = (l_t^{-1} + l_R^{-1})^{-1} + l_{min},
\end{equation}
where $l_t$ is the mean free path due to scattering from TS's,
$l_R$ comes from the Rayleigh scattering, and $l_{min}$ is constant
which must be included to prevent $l$ from becoming unphysically
small. It should be stressed that it is this required cut-off $l_{min}$
that causes the main problem. In fact, it has no natural explanation.
A different situation arises in a-$As$ where the experimental data
were described by a combination of a constant mean free path resulting
from phonon scattering by small holes in samples and the Rayleigh type
scattering. In this case, the Rayleigh scattering becomes important
in the high-frequency limit. Thus, one has to use the interpolation
formula which describes the Rayleigh scattering over the complete
frequency range. It reads~\cite{JTP}
\begin{equation}
\label{eq3}
l_R = D^{-1}\left (\frac{\hbar\omega}{k_B}\right )^{-4} + l_0,
\end{equation}
where D is a constant which has been considered as a fitting
parameter, and $k_B$ is the Boltzman constant.
This approach explains the thermal
conductivity over the full temperature range without the TS scattering.
However, this model is restricted to description of a-$As$
where $\kappa\sim T^3$ has been observed below $0.5 K$
instead of $\kappa\sim T^2$ for the most of glasses.

Let us return to the paper~\cite{freeman} where the following
three characteristic temperature regions have been distinguished:

1. At $T/\Theta\leq 0.01$ the phonon scattering is such that
$l/\lambda\approx 150$ for any type of glass.
Here $\Theta$ is the Debye temperature,
and $\lambda$ is the wavelength of a phonon. The main question is
why this relation is valid for such a varied spectrum of glasses?

2. In the plateau region, $0.01\leq T/\Theta\leq 0.1$,
the mean free path of the phonons rapidly decreases with increasing
$q$. What is the scattering mechanism responsible for this
frequency dependence?

3. At $T/\Theta \geq 0.1$ it is not even clear what excitations are
responsible for thermal transport.

We will try to answer some of these questions below within our model.
First of all, let us reproduce briefly the main results
of the previous paper where the scattering of thermal phonons by WDD
was studied~\cite{osipov}.
For simplicity, in what follows we consider only biaxial WDD
with nonskew axes of rotation (below the term "biaxial WDD" will
refer only to this type of biaxial WDD).

As is known, dislocations and disclinations cause
additional strain fields which lead to the scattering
of phonons. A mean free path of phonons of frequency $\omega$
scattered by the potential associated with a static deformation
of a lattice caused by straight WDD can be calculated within
the generally accepted approach (see., e.g.~\cite{klemens,ziman}).
An effective perturbation energy due to the strain field
caused by a single WDD is written as $U(\vec r)=\hbar\omega\gamma SpE_{AB}$,
where $\hbar\omega$ is the phonon energy with the wavevector $\vec q$,
$\omega=qv$, $\gamma$ is the Gr\"{u}neisen constant, and $E_{AB}$ is
the strain tensor due to WDD.
For a biaxial WDD, the strain matrix is found to have a complex form.
To simplify the problem, we assume that incident phonons lie normally to
disclination lines and choose the suitable geometry with
disclination lines directed along the $z$-axis and
the dipole's arm oriented along the $x$-axis.
In this case, $U(\vec r)$ takes the form:
\begin{equation}
\label{eq4}
U(x,y) = \frac{B}{2}\ln {\frac{(x+L)^2+y^2}{(x-L)^2+y^2}}
\end{equation}
where $B=\hbar qv\gamma\nu (1-2\sigma)/(1-\sigma)$,
$\nu $ is the Frank index, $\sigma$ is the Poisson constant, and
$2L$ is the dipole separation.

In accordance with Eq.(4), the problem reduces
to the two-dimensional case with the phonon mean free path
given by
\begin{equation}
\label{eq5}
l^{-1}_q = n_{i}\int_0^{2\pi}(1-\cos\theta)\Re(\theta)d\theta,
\end{equation}
where $\Re(\theta)$ is an effective differential scattering radius,
and $n_i$ is the areal density of defects.
Within the Born approximation, $\Re(\theta)$ is determined as~\cite{ziman}
\begin{equation}
\label{eq6}
\Re(\theta) =
\frac{qS^2}{2\pi\hbar^2v^2}\overline{|<\vec q|U(\vec r)|\vec q'>|^2},
\end{equation}
where all vectors are two-dimensional ones,
$\vec p=\vec q - \vec q'$, $S$ is a projected area, and
the bar denotes an averaging procedure over
$\alpha$ which defines an angle between $\vec p$ and the $x$-axis.
In other words, it means the averaging over randomly
oriented dipoles in the $xy$ plane.
Evidently, the problem reduces to the estimation of
the matrix element in Eq.(6) with the potential from Eq.(4).
The result is
\begin{equation}
\label{eq7}
U(p, \alpha) =
-\frac{4\pi iB}{p^2}\sin(pL\cos\alpha).
\end{equation}
After averaging of $|U(p,\alpha)|^2$ over $\alpha$
and following integration in Eq.(5) with respect to $\theta$ one obtains
\begin{eqnarray}
\label{eq8}
\l_D^{-1} = 8A^2(\nu L)^2n_iq\Bigl (J_0^2(2qL)+J_1^2(2qL)- \nonumber \\
\frac{1}{2qL}J_0(2qL)J_1(2qL)\Bigr ),
\end{eqnarray}
where $A=\pi\gamma(1-2\sigma)/(1-\sigma)$, and $J_n(t)$ are
the Bessel functions.
Notice that the behavior of $l_{D}$ in Eq.(8)
is actually governed by the only parameter $2L$ which characterizes
the dipole separation.
In the long wavelength limit one gets
$l_D\sim q^{-1}$ while for $\lambda< L$ we obtain that
$l_D\rightarrow const$.
It should be stressed that the appearance of the $q$-independent
region distinguishes remarkably this scatterer from the
others known.

The total phonon mean free path is obtained by combining the
inverse lengths
\begin{equation}
\label{eq9}
l_{tot}(\omega) = (l_R^{-1} + l_{D}^{-1})^{-1}.
\end{equation}
with $l_R$ and $l_{D}$ from Eq.(2) and Eq.(8), respectively.
An important consequence of Eq.(9) is that we do not need any
cut-off procedure in calculating $\kappa$.
\begin{figure}[h]
\epsfxsize=9cm
\centerline{\hspace{7mm} \epsffile{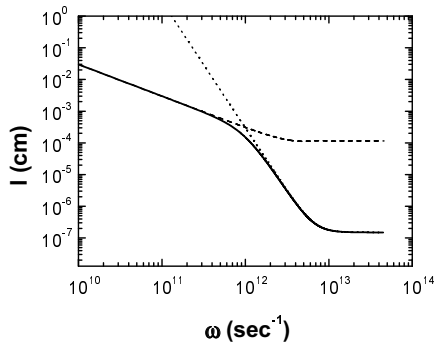}}
\vspace{3mm}
\caption{
Phonon mean free paths $l_D$ (dashed line), $l_R$ (dotted line),
and $l_{tot}$ (solid line) as functions of frequency.
The parameter set used is:
$n_i=5\times 10^{11}$ cm$^{-2}$, $2L=20$\AA, $\nu=0.1$,
$v=4.1\times 10^5$ cm/sec, $A=2.6$, $D=1.0$ cm$^{-1} K^{-4}$,
and $l_0=15$\AA.
}
\label{fig1}
\end{figure}

Fig.1 shows $l_D$, $l_R$, and
$l_{tot}$ for a characteristic set of the model
parameters. Notice that the parameters for the Rayleigh scattering
are chosen to be in agreement with those for a-$SiO_2$~\cite{JTP}.
One can clearly see that at low frequencies the dominant
scattering mechanism comes from the WDD-induced scattering,
then there appears a region with the two scattering processes
involved, and finally the Rayleigh scattering becomes dominant at high
frequencies. This specific behavior results from
the fact that both $l_R$ and $l_D$ tend to constants
when $q$ increases. Obviously, any $q$ dependence of scatterers
would break this picture at large $q$. The main parameters
which actually define $l_{tot}$ are $D$ from the Rayleigh scattering
and $2L$ from the WDD scattering.
\begin{figure}[h]
\epsfxsize=9cm
\centerline{\hspace{7mm} \epsffile{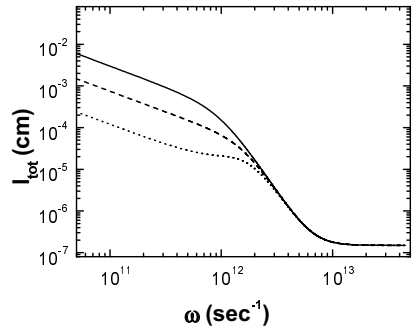}}
\vspace{3mm}
\caption{
Phonon mean free path $l_{tot}$ as a function of frequency
at $2L=20$\AA\  (solid line), $2L=40$\AA\  (dashed line), and
$2L=100$\AA\  (dotted line).
The parameter set is the same as in Fig.1.
}
\label{fig2}
\end{figure}
As an example, Fig.2 shows $l_{tot}$ for different dipole separations.
It is interesting to note that the size of the transition region
decreases remarkably with $2L$ increasing.
Notice that this region alone is responsible for the plateau in the thermal
conductivity.
\begin{figure}[h]
\epsfxsize=9cm
\centerline{\hspace{7mm} \epsffile{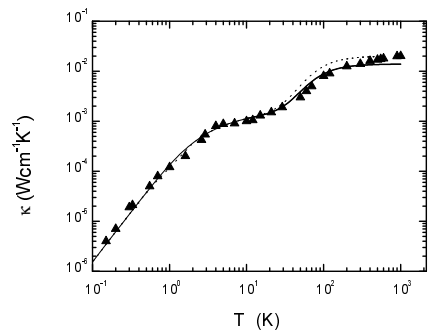}}
\vspace{3mm}
\caption{
Thermal conductivity vs temperature calculated according to
Eq.(10) with $l_{tot}$ from Eq.(9) at $l_0=10$\AA\  (solid line)
and $l_0=15$\AA\  (dotted line). The parameter set is the
same as in Fig.1, $\Theta=342 K$.
Experimental data for a-$SiO_2$ from Ref.[1]
are indicated by triangles.
}
\label{fig3}
\end{figure}
The last step is to calculate $\kappa$ in Eq.(1) with $l_{tot}$.
For this purpose, it is convenient to rewrite Eq.(1) in the
dimensionless form
\begin{equation}
\label{eq10}
\kappa = \frac{k_B^4T^3}{2\pi^2\hbar^3v^2}
\int_0^{\Theta/T}x^4e^x(e^x-1)^{-2}l_{tot}(x)dx,
\end{equation}
where $x=\hbar\omega/k_BT$, and
the specific heat capacity is chosen in the standard Debye form.
The result is shown in Fig.3 where the experimental points
for a-$SiO_2$ are marked by triangles. As is seen, there is a good
agreement over a wide temperature range. For $l_0=15$\AA\
some deviations from the experimental data appear
above $40 K$ but the correct qualitative behavior retains even
at high temperatures.
The best fit for a-$SiO_2$ is obtained for $2L=20$\AA. Notice that
this value agrees with a cluster size expected in vitreous
silica~\cite{JCP}.
Let us stress once more that we have no need for poorly defined
parameters like the cut-off $l_{min}$ in Eq.(2) or an additional
to $l_R$ constant free path proposed in~\cite{JTP} to fit the data
below $0.5 K$.

Now we can try to answer the questions formulated in the beginning
of this paper. In accordance with Fig.1, at low frequencies
the dominant contribution comes from $l_D$. One can easily estimate
the value of $l_D/\lambda$ in this region. Spreading
out Eq.(8) at $qL\ll 1$ one gets
\begin{equation}
\label{eq11}
\frac{l_D}{\lambda} = \frac{1}{8\pi A^2(\nu L)^2n_i}.
\end{equation}
This is the constant which depends on the model parameters.
Thus, in accordance with our model, the only parameters
which determine the behavior of glasses at low temperatures
come from their microstructure which includes biaxial WDD
as the basic elements. The main parameters are the density of WDD,
their Frank index, and the dipole separation.
In particular, for our choice of parameters, $l_D/\lambda\sim 135$.
Let us emphasize that the experimental
data for various glasses given
in~\cite{zeller} can be described by minor variations of
these main parameters (together with $D$).
Thus the relation $l_{tot}/\lambda\sim 150$ is indeed satisfied
for all these dielectric glasses.
Based on the results of our analysis, we can conclude that:

\noindent
(i) the WDD-induced scattering of phonons can explain the established
relation $l/\lambda\sim 150$ for various glasses;

\noindent
(ii) in the plateau region both mechanisms (WDD-induced and
the Rayleigh scattering) should be accounted for to produce a necessary
behavior;

\noindent
(iii) with increasing temperature, the Rayleigh scattering becomes
dominant. While there are some deviations from the experimental data
at high temperatures, the general behavior looks quite correct.
Thus nothing else but phonons transport the heat at $T>0.1\Theta$,
and no any other excitations are needed.

To summarize, we have presented a new look at the problem
of the thermal conductivity in amorphous dielectrics.
The main suggestion is based on the introduction of biaxial
WDD as the basic structural elements of glasses.
The unique properties of the WDD-induced phonon scattering
enabled us to describe the universal behavior of the thermal
conductivity of amorphous dielectrics over a wide temperature range.
Thus, the results obtained support the cluster picture of amorphous
dielectrics suggested by Phillips~\cite{JCP} and shed some light
on the relationship between the microstructure of glasses and
their transport properties.

\vskip 0.5cm

This work has been supported by the Russian Foundation for Basic
Research under grant No. 97-02-16623.

\end{document}